\documentclass[12pt]{article}

\usepackage{graphics}
\usepackage{graphicx}
\usepackage{subcaption} 
\usepackage{color}
\usepackage{amssymb}
\usepackage{amsmath}
\usepackage{verbatim}

\usepackage[colorlinks = true,
linkcolor = blue,
urlcolor  = blue,
citecolor = blue,
anchorcolor = blue]{hyperref}
\usepackage[numbers,square,sort&compress]{natbib}

\usepackage{authblk}
\usepackage{cancel}

\numberwithin{equation}{section}
\begin{document}
	\title{\bf Curved Corner Contribution to the Entanglement Entropy in an Anisotropic Spacetime} 
	\author[1]{Mostafa Ghasemi}
	\author[2]{Shahrokh Parvizi}
	\affil[1]{ School of Particles and Accelerators, Institute for Research in Fundamental Sciences (IPM)\\
		P.O. Box 19395-5531, Tehran, Iran}
	\affil[2]{ Department of Physics, School of Sciences,
		Tarbiat Modares University, P.O.Box 14155-4838, Tehran, Iran} 
	\affil[ ]{\textit {Email: \href{mailto:ghasemi.mg@ipm.ir}{ghasemi.mg@ipm.ir},   \href{mailto:parvizi@modares.ac.ir}{parvizi@modares.ac.ir}}}
	\date{\today}                     
	\setcounter{Maxaffil}{0}
	\renewcommand\Affilfont{\itshape\small}
	
	\maketitle

	\abstract 

In this article, we explore the divergences and universal terms of the holographic entanglement entropy for singular regions in anisotropic and nonconformal theories that are holographically dual to geometries with a hyperscaling violation, parameterized by two parameters $z$ and $\theta$.  We study a curved corner in anisotropic space with arbitrary $\theta$ and $z$. We choose the region to be shape invariant under the scaling of spacetime. For this case, we show that the contribution of the singularity to the entanglement entropy depends on $z$ and $\theta$ values. We identify the structure of various divergences that may appear, especially those which give rise to a universal contribution in the form of logarithmic or double logarithmic terms. In the range $z>1$, for values $z=2k/(2k-1)$ with some integer $k$ and $\theta=0$, Lifshitz geometry, we find a double logarithmic term. In the range $z<0$, for values $\theta=1-2n|z-1|$ with some integer $n$ we find a logarithmic term.



\section{Introduction} \label{intro}
The gauge/gravity duality \cite{Ref1,Ref2,Ref3} provides a framework to explore various aspects of the strongly coupled gauge field theories. By this duality, the $(d+1)$-dimensional field theory duals to a gravitational theory in a $(d+2)$-dimensional spacetime. Although 
the early studies in this context lie on investigating duality between a conformal field theory ($CFT$) on the boundary and the gravity on the $AdS$ background, over the years, it was generalized to account for a diverse range of quantum systems \cite{Ref4, Ref5, Ref6, Ref7, Ref8, Ref9, Ref10, Ref11, Ref12}.

One of the extension is theories which are scale invariant but not conformal. These are corresponding to the Lifshitz fixed point and characterized by a dynamical critical exponent $z$. The gravity dual of these theories is defined in the following background metric \cite{Ref13,Ref14},
\begin{equation}
ds ^{2}= \frac{L^{2}}{r^{2}}(-r^{-2(z-1)}dt^{2}+dr^{2}+d\vec{x}_{d}^{2}),
\end{equation}
where $L$ is the $AdS$ curvature scale. The following scale transformation leaves the metric invariant,
\begin{equation} 
t\rightarrow \lambda^{z}t, \qquad r\rightarrow \lambda r \qquad x_{i}\rightarrow \lambda x_{i}, \qquad ds ^{2}\rightarrow  ds ^{2}.
\end{equation}

Another extension, is the strongly coupled field theories which are dual to geometries characterized by two parameters, a dynamical exponent $z$, and a hyperscaling violation exponent $\theta$. The metric of these geometries is defined as \cite{Ref15}
\begin{equation}\label{metric2}
ds ^{2}= \frac{L^{2}}{r^{2}}(\frac{r}{r_{F}})^{\frac{2\theta}{d}}(-r^{-2(z-1)}dt^{2}+dr^{2}+d\vec{x}_{d}^{2}),
\end{equation}
where $r_{F}$ represents the scale below which the above metric provides the proper gravitational dual for strongly coupled theory \cite{Ref22}. The scaling transformations of the coordinates are defined as 
\begin{equation}\label{scaling1}
t\rightarrow \lambda^{z}t, \qquad r\rightarrow \lambda r \qquad x_{i}\rightarrow \lambda x_{i}, \qquad ds ^{2}\rightarrow \lambda^{\frac{2\theta}{d}} ds ^{2}
\end{equation}
The metric \eqref{metric2} transforms covariantly under these transformations and is called hyperscaling violating Lifshitz metric (hvLf). In the context of holography, these metrics have been extensively studied \cite{Ref16,Ref17,Ref18,Ref19,Ref20,Ref21,Ref22,Ref23,Ref24}.

As an application of the AdS/CFT correspondence, one can evaluate the entanglement entropy of the boundary theory by Ryu-Takayanagi (RT) prescription \cite{Ref25,Ref26}. The entanglement entropy is one of the important measures of entanglement feature in quantum systems and emerges in diverse research areas \cite{Ref11,Ref12,Ref27,Ref28,Ref29,Ref30,Ref31,Ref32,Ref33,Ref34,Ref35,Ref36,Ref37}. In the context of quantum field theory, the entanglement entropy of a sub region ${A}$ is defined as $ S=-Tr( \rho_{A} log\rho_{A})$, where the reduced density matrix $ \rho_{A}$ is obtained by tracing out the degrees of freedom of complementary region $\bar{A}$ of $ {A}$, $\rho_{A}=Tr_{\bar{A}}(\rho)$. 

In general, the entanglement entropy is $UV$ divergent due to the short-range correlations across the so-called entangling surface, the boundary of two regions. So to have a well-defined quantity it must be regularized. This regularization depends on the geometry and topology of the background as well as the entangling surface. As an example,
in the vacuum of a $(2+1)$- dimensional $CFT$, the entanglement entropy for a smooth entangling surface takes the following form
\begin{equation}
S_{EE}=\beta \frac{l_\Sigma}{\delta} -F,
\end{equation}
where $l_\Sigma$ is the length of the entangling surface, $\delta$ is a $UV$ cut-off, and $\beta$ is a scheme-dependent coefficient that depends on the details of the underlying theory. The leading term exhibits the``area law" \cite{Ref38,Ref39} and the second term $F$ is a universal term independent of the regularization scheme.
On the other hand, for a corner which is singular in the entangling surface, the entanglement entropy contains a universal contribution,
\begin{equation}\label{SEE-aOmega}
S_{EE}=\alpha\frac{H}{\delta}- a(\Omega)\log(\frac{H}{\delta})+O(1),
\end{equation}
where $\Omega$ is the opening angle, $\alpha$ is the scheme dependent constant, and $H$ denotes the characteristic length of the entangling surface. $a(\Omega)$ is a coefficient of the new logarithmic term that appears owing to the singularity of entangling surface and gives the universal contribution to the $EE$. The Lorentz invariance and subadditivity properties of the entanglement entropy require that $a(\Omega)$ to be a positive convex function \cite{Ref40,Ref41,Ref42,Ref43,Ref44,Ref45,Ref46},
\begin{align}\label{a-properties}
a(\Omega)\geq 0, \qquad a'(\Omega)\leq 0, \qquad a''(\Omega)\geq 0.
\end{align}
Moreover, for $\Omega=\pi$, the corner goes to a smooth half plane and we expect $a(\pi)=0$. On the other hand, for a pure state, the entanglement entropy of any region $V$and its complement $\overline{V}$ are equal, $S(V)=S(\overline{V})$. This implies that $a(\Omega)$ should have a Taylor series expansion with even powers of $(\pi-\Omega)$ as $\Omega\rightarrow \pi$.

A similar universal contribution appears in the entanglement entropy of conformal field theories,
for which depending on the dimension of space time and singularity of entangling surface, 
logarithmic or double logarithmic terms appear \cite{Ref47,Ref48,Ref49,Ref50,Ref51,Ref52,Ref53}.

There are similar stories on the EE of the relevant perturbation of  conformal field theories which induces a universal logarithmic \cite{Ref54,Ref55,Ref56,Ref57,Ref58,Ref59,Ref60,Ref61,Ref62,Ref63,Ref64} or double logarithmic term in the entanglement entropy \cite{Ref65}.
The importance of these universal terms is that their coefficients are scheme independent and encode the universal data of the underlying quantum field theory \cite{Ref44}.

In the context of non-conformal theories, there are similar results.
In the case of hyperscaling violation background, typical computations in $d=3$ \cite{Ref66} shows that for $\theta=1$, the entanglement entropy shows a logarithmic violation of the area law, and corresponds to the Fermi surface in the boundary theory \cite{Ref17,Ref22,Ref23}. The log terms will appear by adding terms with higher derivatives to the Einstein gravity or higher-dimensional space-times \cite{Ref67,Ref68}. For other aspects of this context see  \cite{Mishra:2016yor,Mishra:2018tzj,MohammadiMozaffar:2017nri,MohammadiMozaffar:2017chk}.

Motivated with recent work \cite{Giataganas:2017koz} in which the scaling solutions in IR limit are found, we examine the entanglement structure of the anisotropic strongly coupled field theories for which a gravitational theory dual can be defined by the Einstein-axion-dilaton action 
\begin{align}\label{axion-dilaton}
S&=\frac{1}{2\ell_p^3}\int d^5x\sqrt{-g}\Big(R+\mathcal{L}_M\Big) \nonumber\\
\mathcal{L}&=-\frac{1}{2}(\partial\phi)^2+V(\phi)-\frac{1}{2}Z(\phi)(\partial\chi)^2 
\end{align}
in which $V$ is a generic potential for the dilaton field $\phi$, and $Z$ is an arbitrary coupling between axion $\chi$ and dilaton fields. With respect to generic choice of the $V$ and $Z$ we can reach to various solutions. We consider the following case \cite{Giataganas:2017koz,Giataganas:2018ekx},
\begin{align}
V(\phi)&=6e^{\sigma\phi}, \qquad Z(\phi)=e^{2\gamma\phi} 
\end{align}
The solution of equations of motion generated by \eqref{axion-dilaton} is a Lifshitz-like anisotropic hyperscaling violation metric which exhibits the arbitrary critical exponent $z$ and a hyperscaling violation exponent $\theta$, both are functions of constants $\sigma$ and $\gamma$. The metric of dual geometry is defined as\footnote{In order to absorb constants we rescaled the coordinates of \cite{Giataganas:2017koz}.} 
\begin{equation}\label{metric1}
ds ^{2}= L^{2}r^{\frac{2\theta}{dz}}(\frac{-dt^{2}+dr^{2}+d\vec{x}_{d-1}^{2}}{r^{2}}+\frac{dy^{2}}{r^{\frac{2}{z}}}).
\end{equation}
Noting that this solution is an exact metric along UV to IR.	
It has a scaling property defined as 
\begin{equation}\label{scaling}
t\rightarrow \lambda t, \qquad r\rightarrow \lambda r \qquad x_{i}\rightarrow \lambda x_{i}, \qquad y\rightarrow \lambda^{\frac{1}{z}} y, \qquad ds ^{2}\rightarrow \lambda^{\frac{2\theta}{dz}} ds ^{2},
\end{equation}
where the metric transforms covariantly under these transformations.
The Lagrangian setup for these kind of solutions is presented in  \cite{Giataganas:2018ekx}, while the exact scaling factors of background solution is in \cite{Giataganas:2018rbq}.\footnote{We would like to thank Dimitrios Giataganas for pointing this out to us.} 
There are other related works in this background and context \cite{Giataganas:2018rbq,Gursoy:2018ydr,Ghasemi:2019xrl}.

Note also that the anisotropic metric \eqref{metric1} can be understood as a double wick rotation of the standard Lifshitz/HsV metrics, e.g. \eqref{metric2}, as well as scaling of $z$ and $\theta$ \cite{deBoer:2011wk,Alishahiha:2012cm}.

In this paper, we explore the effects of singularity as well as anisotropy in spatial direction on the entanglement entropy. Our motivation is to find the universal terms just like in the case of  corner contribution in the 2+1-dimensional isotropic space-time.

It is worth to note that the parameters $z$ and $\theta$ are constrained by the null energy condition which is 
\begin{align}
\xi^\mu\xi^\nu G_{\mu\nu} \geq 0,
\end{align}
where $\xi^\mu$ is any null vector and $G_{\mu\nu}$ is the Einstein tensor. We adopt the null vector to be
\begin{align}
(\xi^t,\xi^r,\xi^x,\xi^y)=\frac{1}{L}(\sqrt{a^2+b^2+c^2}r^{1-\frac{\theta}{d z}},ar^{1-\frac{\theta}{d z}},br^{1-\frac{\theta}{d z}},cr^{\frac{1}{z}-\frac{\theta}{d z}}).
\end{align}
with $d=2$. 
Then the null energy condition (NEC) implies two independent inequalities as follows,
\begin{align}\label{NEC1}
\theta^2+2z(1-\theta)-2 &\geq 0 ,\\
\label{NEC2}
(z-1)(1+2z-\theta) &\geq 0.
\end{align} 

Our goal in this paper is to calculate the entanglement entropy of the curved kink region in the \eqref{metric1} background via RT prescription. According to it, the holographic entanglement entropy of a sub-region $V$ on the boundary theory is given by
\begin{equation}
S_{EE}= \frac{Area(\Sigma)}{4G_{N}},
\end{equation}
in which $\Sigma$ is the bulk minimal surface which is homologous to $V$ and $\partial_{\Sigma}$ matches the entangling surface $\partial_{V}$ on the boundary. We then investigate the divergent structure of the entanglement entropy and look for any universal term.

Our paper is organized as follows. In section \ref{sec:HEE-curved}, we study a shape invariant singular region with orientation along the anisotropic scaling direction $y$. By shape invariant, we mean a curved corner region with boundary at $y=\pm a x^{1/z}$ which is invariant under the scaling \eqref{scaling}. This choice, practically, simplifies our equations a lot. 
In section \ref{sec:rotating}, we comment on the singular surface lying along isotropic direction $x$. In the last section, we summarize and discuss our results. Some lengthy formulas and calculations are given in the appendix \ref{app1}. 


\section{Holographic entanglement entropy of a curved kink}
\label{sec:HEE-curved}

\begin{figure}
	\captionsetup{width=0.8\textwidth}
	\begin{center}
		\includegraphics[height=55mm]{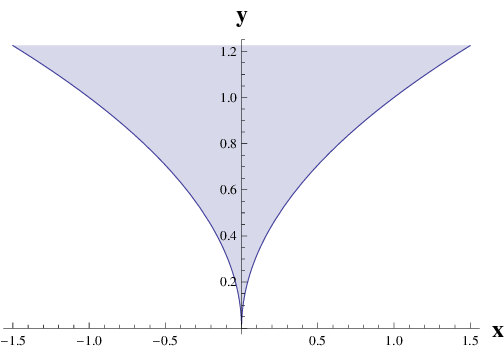} \;\;
		\includegraphics[height=55mm]{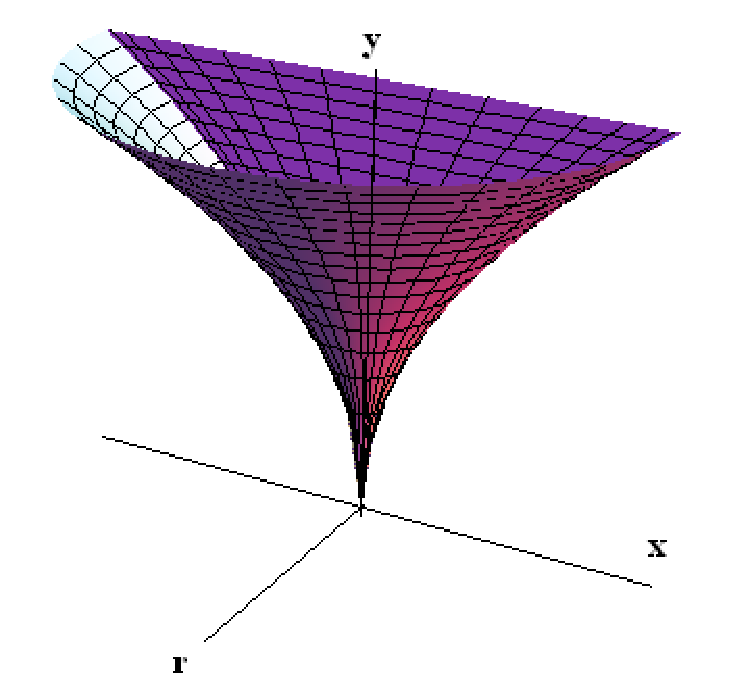} \\
		\caption{The left panel shows the entangled region in the $xy$ plane with $b=1/2$ and $a=3/2$. The right panel illustrates the corresponding bulk minimal surface.}
		\label{fig:RT-surface}	
	\end{center}
\end{figure}

\begin{figure}
	\captionsetup{width=0.8\textwidth}
	\begin{center}
		\includegraphics[height=30mm]{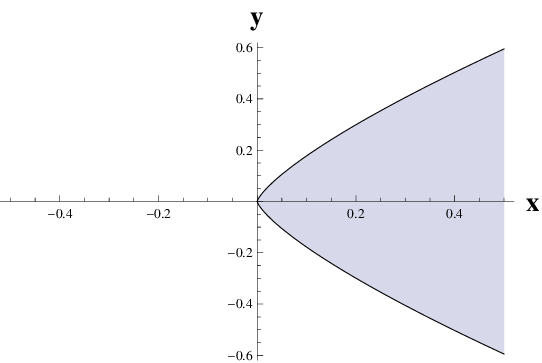} \;\;
		\includegraphics[height=30mm]{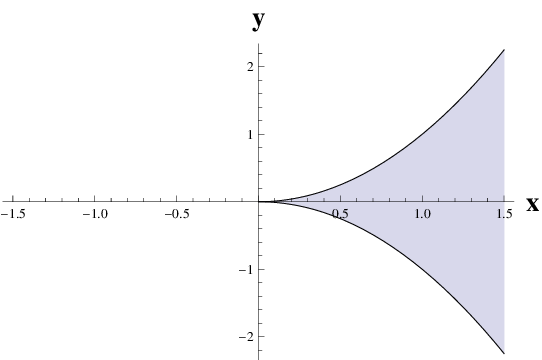} \;\;
		\includegraphics[height=30mm]{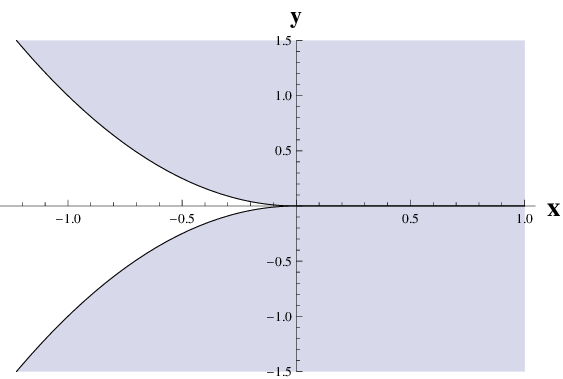} \\
		i \hspace{45mm} ii \hspace{45mm} iii
		\caption{The entangled region in the $xy$ plane. i) $b<1$ and $0<\varphi_0<\pi/2$, ii) $b>1$ and $0<\varphi_0<\pi/2$, iii) $b>1$ and $\pi/2<\varphi_0<\pi$. For a pure state, $S(\text{ii})=S(\text{iii})$. }	
		\label{fig:regions}
	\end{center}
\end{figure}

In this section, we study the entanglement entropy of a curved kink region given in the time slice $t_{E}= 0 $ as  $V=\{|y|\leq a x^{b} ,\;\; 0\leq x \leq \tilde{H}\}$, where $\tilde{H}$ is an $IR$ cut-off, $a$ is a constant which represents the wideness of the curved Kink and $b$ is an arbitrary constant. See Fig. \ref{fig:RT-surface}.
\\
We choose the metric
\begin{equation} 
ds ^{2}= L^{2} r^{\frac{\theta}{z}}(\frac{-dt^{2}+dr^{2}+dx^{2}}{r^{2}}+\frac{dy^{2}}{r^{\frac{2}{z}}}),
\end{equation}
and define our coordinates as
\begin{equation}
x=\rho\cos \varphi, \qquad      y=\rho^{b}\sin^{b} \varphi
\end{equation}
In this coordinate, the entangled region is $V=\{-\varphi_0 \leq \varphi \leq \varphi_0 ,\;\; 0\leq \rho \leq H\}$
where $H=\tilde{H}/\cos\varphi_0$ and we define $\varphi_{0}$ as $a=\tan^b\varphi_{0}$ which introduces the boundary of region and consider $0 \leq \varphi_{0} \leq \pi$.
Different regions are illustrated in Fig. \ref{fig:regions}.
\\

We parametrize the bulk minimal surface as $r=r(\rho,\varphi)$ and the induced metric on the bulk minimal surface becomes
\begin{align}
ds^{2}&=\gamma_{ij}dx^idx^j \nonumber\\
&=L^{2}r^{\frac{\theta-2}{z}}\Big[\Big(b^2\rho^{2b-2}\sin^{2}\varphi +\frac{r'^{2}+ \cos^2\varphi}{r^{2-\frac{2}{z}}}\Big) d\rho^{2} +\Big(\rho^{2b}\cos^{2}\varphi +\frac{\dot{r}^{2}+ \rho^2\sin^2\varphi}{r^{2-\frac{2}{z}}}\Big)d\varphi^{2} \nonumber\\
& +\Big(b\rho^{2b-1}\sin(2\varphi) +\frac{2\dot{r}r'-\rho\sin(2\varphi)}{ r^{2-\frac{2}{z}}}\Big)d\rho d\varphi
\Big],
\end{align}
where $\dot{r}= \partial_{\varphi}r$, and $r'=\partial_{\rho}r$.
So we find that
\begin{align}
\sqrt{\gamma}&=L^{2}r^{\frac{\theta}{z}-2}
\Big[(r'\rho\sin \varphi+\dot{r}\cos \varphi)^{2}\nonumber\\
&+r^{2-\frac{2}{z}}\rho^{2b-2}\Big(\frac{\rho^2}{4}(1+b-(b-1)\cos(2\varphi))^2+(b\dot{r}\sin \varphi-r'\rho\cos\varphi)^{2}\Big)\Big]^{\frac{1}{2}}.
\end{align}
The holographic entanglement entropy is given by
\begin{equation}\label{SEE-1}
S_{EE}=\frac{1}{4G_{N}}\int d\rho d\varphi\sqrt{\gamma} =\frac{L^{2}}{2G_{N}}\int_{\rho_{m}}^{H}d\rho\int_{0}^{\varphi_0-\epsilon}
d\varphi\mathcal{L},
\end{equation}
in which $\delta$ is a $UV$ cut-off, $\epsilon$ is an angular cut-off, $\rho_m$ depends on $\delta$ and is defined later and $\mathcal{L}$ is given by
\begin{align}\label{Lagr}
\mathcal{L}=\sqrt{\gamma}/L^{2}. 
\end{align}

The equation of motion is given in the appendix \ref{app1} and is solved for an important special case\footnote{Another interesting case is a curved corner in the symmetric space, i.e. $\theta=0$, $z=1$ with an arbitrary $b$. This was analyzed in \cite{Bueno:2019mex} by another approach.} where $\theta$ and $z$ are arbitrary while we set $b=1/z$. This simplifies the EoM, and is shape invariant under the scaling \eqref{scaling} which means that $a$ remains fixed. We find the near boundary solution which contributes to the singular part of the entanglement entropy.

\subsection{Shape invariant curved region in an asymmetric space}\label{subsec:shape-invariant}
In this subsection, we consider asymmetric hyperviolating space-time with arbitrary $\theta$ and $z$. The entangled region is chosen to be a curved corner with $b=1/z$. This choice has two advantages. Firstly, it simplifies the EoM to a nonlinear ordinary differential equation and secondly, it is shape invariant under scaling \eqref{scaling}, i.e., $a=\tan^{1/z}\varphi_{0}$ remains intact. The EoM and its solutions are given in the appendix \ref{app1}. In this case, we take the ansatz  $r=\rho h^z(\varphi)$, so that the near boundary limit corresponds to small $h$. Now we change the independent variables from $(\rho,\varphi)$ to $(r,h)$, such that $\varphi=\varphi(h)$. Then we find the entanglement entropy to be,
\begin{equation}\label{eq 11}
S_{z}=\frac{L^{2}}{2G_{N}}\frac{1}{2z}\int_{\delta}^{r_{m}}\frac{dr}{r^{1-\frac{\theta}{z}}}\int_{h_{0}}^{h_{c}} dh\mathcal{L}_h 
\end{equation}
in which we introduced $r=\delta$ as a UV cut-off, and we have changed the integration variable from $\varphi$ to $h$, and $\mathcal{L}_h$ is  defined as
\begin{align}\label{int 1}
\mathcal{L}_h= \frac{2\sqrt{h^{2}z^{2}(h\sin \varphi\dot{\varphi}+z\cos \varphi)^{2}+h^{2}(\sin^2\varphi+z\cos^2\varphi)^{2}\dot{\varphi}^{2}+z^{2}h^{2z}(h\cos \varphi\dot{\varphi}-\sin \varphi)^{2}}}{h^{2+z}}
\end{align}
The equation of motion for $\varphi(h)$ is given in the appendix \ref{app1} and is solved perturbatively in terms of small $h$ near the boundary, with boundary condition $\varphi(0)=\varphi_{0}$ at $h=0$. Let us consider different $z$ ranges as follows.

\subsubsection{$z>1$}\label{zgt1}
For $z>1$, we find the following expansion as a solution to \eqref{EoM-phi},
\begin{align}
&\varphi=\varphi_0+\sum_{n=1}\varphi_{2n}(h^{2})^{n(z-1)+1}
\nonumber\\
&\quad=\varphi_0+\varphi_{2}h^{2z}+\varphi_{4}(h^{2})^{2z-1}+\varphi_{6}(h^{2})^{3z-2}+\cdots,
\end{align}
where some $\varphi_{2n}$'s are derived as,
\begin{align}
\varphi_2&=\frac{(z-1) \tan \varphi_0}{(\theta+z-2) ((z-1) \cos (2\varphi_0)+z+1)},\nonumber\\
\varphi_4&=-\frac{(\theta-1) \tan ^2\varphi_0}{z (2 z-1) (\theta+3 z-4)}\varphi_2 \nonumber
\end{align}

Near the boundary, assuming $z>1$, the integrand of \eqref{int 1} behaves as
\begin{align}
\mathcal{L}_h \sim \frac{2z^{2}\cos \varphi_{0}}{h^{z+1}} \sqrt{1+\frac{\tan^2\varphi_0}{z^2}h^{2(z-1)}}+\cdots
\end{align}
where dots denotes nonsingular contributions. It follows then
\begin{align}\label{eq 2}
\mathcal{L}_h &\sim \sum_{n=0}a_n h^{2n(z-1)-(z+1)}+\cdots,\nonumber\\
&\sim \frac{2z^2\cos \varphi_{0}}{h^{z+1}} + \frac{\sin \varphi_{0}\tan\varphi_{0}}{h^{3-z}}- \frac{\sin \varphi_{0}\tan^3\varphi_{0}}{4z^2h^{3z-5}}+\frac{\sin \varphi_{0}\tan^5\varphi_{0}}{8z^4h^{5z-7}} + \cdots,
\end{align}
in which $a_n\sim\sin\varphi_0\tan^{2n-1}\varphi_0$. In \eqref{eq 2},  
the singular terms are restricted to $n\leq n_{max}$ which is the greatest integer satisfying $n_{max}<(z+1)/(2(z-1))$.

Now by using the relations \eqref{eq 2} we can isolate the divergent part of integral \eqref{eq 11}, and make it finite
\begin{align}\label{SEE2}
S_{EE}&=\frac{L^{2}}{2G_{N}}\frac{1}{2z}\int_{\delta}^{r_{m}}\frac{dr}{r^{1-\frac{\theta}{z}}}\int_{h_{0}}^{h_{c}} dh\mathcal{L}_h , \nonumber\\
&=\frac{L^{2}}{2G_{N}}\frac{1}{2z}(I_{1}+I_{2})
\end{align}
where $h_{c}(\delta)=(\frac{\delta}{H})^{\frac{1}{z}}$. $I_{1}$ and $I_{2}$ are introduced by subtracting and adding singular terms as follows
\begin{align}
I_{1}&=\int_{\delta}^{r_{m}}\frac{dr}{r^{1-\frac{\theta}{z}}}\int_{h_{0}}^{h_{c}} dh
\Big[\mathcal{L}_h  - \sum_{n=0}a_n h^{2n(z-1)-(z+1)}\Big] \nonumber\\
\label{I2}
I_{2}&=\int_{\delta}^{r_{m}}\frac{dr}{r^{1-\frac{\theta}{z}}}\int_{h_{0}}^{h_{c}} dh
\Big(\sum_{n=0}a_n h^{2n(z-1)-(z+1)}\Big) 
\end{align}
We then differentiate $I_1$ and $I_2$ with respect to $UV$ cut-off $\delta$ and look for various divergent terms. One finds
\begin{align}
\frac{dI_{1}}{d\delta}&=-\frac{1}{\delta^{1-\frac{\theta}{z}}}\int_{h_{0}}^{0} dh
\Big[\mathcal{L}_h - \frac{2z^{2}\cos \varphi_{0}}{h^{z+1}} - \frac{\sin \varphi_{0}\tan\varphi_{0}}{h^{3-z}} +\cdots \Big] \,,\\
\label{dI2ddelta}
\frac{dI_{2}}{d\delta}&= \frac{F(h_{0})}{\delta^{1-\frac{\theta}{z}}}- \frac{a_k}{z\delta^{1-\frac{\theta}{z}}}\log\frac{\delta}{H}+\sum_{n}\frac{a_n H^{1-2n(1-\frac{1}{z})}}{z-2n(z-1)}\delta^{\frac{\theta}{z}-2+2n(1-\frac{1}{z})} \nonumber\\
&= \frac{F(h_{0})}{\delta^{1-\frac{\theta}{z}}}- \frac{a_k}{z\delta^{1-\frac{\theta}{z}}}\log\frac{\delta}{H}+2z\frac{H\cos \varphi_{0}}{\delta^{2-\frac{\theta}{z}}}  +\frac{\sin \varphi_{0}\tan\varphi_{0}}{2-z} \frac{H^{\frac{2-z}{z}}}{\delta^{\frac{2-\theta}{z}}}+\cdots,
\end{align}
in which the logarithmic term appears from the integration of $k$-th term in \eqref{I2} only when $z=2k/(2k-1)$.

In the final result, at some special values of $z$ and $\theta$ the logarithmic term appears. The first important case is at $\theta=0$,
\begin{align}
S_{EE}=\frac{L^{2}}{2G_{N}}\frac{1}{2z}\Big[&\Big(-\int_{h_{0}}^{0} dh
\Big[\mathcal{L}_h  - \sum_{n=0}^{n_{max}}a_n h^{2n(z-1)-(z+1)}\Big]+F(h_0)
\Big)\log (\frac{\delta}{H})\nonumber\\
&+\frac{a_k}{2z}\log^2 (\frac{\delta}{H})+\cdots\Big],
\end{align} 
where the double log term appears at $\theta=0$ and $z=2k/(2k-1)$ for some positive integer $k$. These values of $\theta$ and $z$ are consistent with the null energy conditions \eqref{NEC1} and \eqref{NEC2}. Noting that, in this case the universal term which is independent of the regularization scheme, is the double log term \cite{Ref51, Ref65}.

Other log terms appear when there is a term $1/\delta$ in the expansion of $dI_2/d\delta$ in \eqref{dI2ddelta}. It happens when
\begin{align}\label{thetam}
\theta= z-2m(z-1)
\end{align}
for some integer $0\leq m \leq n_{max}$. This again satisfies the NEC inequalities. Then one finds,
\begin{align}\label{log1}
&S_{EE}=\frac{L^{2}}{2G_{N}}\frac{1}{2z}\Big[\frac{a_mH^{\theta/z}}{\theta}\log (\frac{\delta}{H})+\cdots\Big].
\end{align}
We collect all terms as 
\begin{align}\label{total-zgt1}
S_{EE}^{(y,z>1)}&=\frac{L^{2}}{2G_{N}}\frac{1}{2z}\Big[\Big(F(h_0)-\int_{h_{0}}^{0} dh
(\mathcal{L}_h  - \sum_{n=0}^{n_{max}}a_n h^{2n(z-1)-(z+1)})
\Big)\Big(\log \frac{\delta}{H}+\frac{z}{\theta}\delta^{\theta/z}\Big)\nonumber\\
&+\frac{a_k}{2z}\log^2 (\frac{\delta}{H})+\frac{a_mH^{\theta/z}}{\theta}\log (\frac{\delta}{H}) \nonumber\\
&+\sum_{n=0}^{n_{max}}\frac{a_n z H^{\theta/z} }{(z-2 n (z-1)) (2 n (z-1)+\theta-z)}\Big(\frac{\delta}{H}\Big)^{\frac{2 n (z-1)+\theta-z}{z}}+\cdots\Big].
\end{align}
where the superscript $y$ in $S_{EE}^{(y,z>1)}$ denotes that the region bisector is along the $y$-axis.
The double log term appears when $\theta=0$ and $z=2k/(2k-1)$ for some integer $k$ and the log term appears when $\theta= z-2m(z-1)$ for some integer $m$. Note that these two terms do not appear simultaneously.



\subsubsection{$0<z<1$}\label{zlt1}
Here, from \eqref{EoM-phi} we find,
\begin{align}
&\varphi=\varphi_0+\varphi_{2}h^{2}+\varphi_{4}h^4+\tilde{\varphi}_1h^{4-2z}+\tilde{\varphi}_{3}h^{6-4z}+\cdots,
\end{align}
Again we will see that only $\varphi_0$ contributes to the singular part of integrand \eqref{int 1} which near the boundary can be expanded as
\begin{align}
\mathcal{L}_h \sim \frac{2z\sin \varphi_{0}}{h^{2}} \sqrt{1+z^2h^{2(1-z)}\cot^2\varphi_0}+\cdots
\end{align}
then
\begin{align}\label{L-zlt1}
\mathcal{L}_h &\sim \sum_{n=0}b_n h^{2n(1-z)-2}+\cdots,\nonumber\\
&\sim \frac{2z\sin\varphi_{0}}{h^{2}} + \frac{z^2\cos \varphi_{0}\cot\varphi_{0}}{2h^{2z}}- \frac{z^4\cos \varphi_{0}\cot^3\varphi_{0}}{8h^{4z-2}}+ \cdots.
\end{align}
and the final result is
\begin{align}\label{total-zlt1}
S_{EE}^{(y,0<z<1)}&=\frac{L^{2}}{2G_{N}}\frac{1}{2z}\Big[\Big(\tilde{F}(h_0)-\int_{h_{0}}^{0} dh
(\mathcal{L}_h  - \sum_{n=0}^{n_{max}}b_n h^{2n(1-z)-2})
\Big)\frac{z}{\theta}\delta^{\theta/z}\nonumber\\
&+\frac{b_\ell H^{\theta/z}}{\theta}\log (\frac{\delta}{H})
\nonumber\\
&+\sum_{n=0}^{n_{max}}\frac{b_n zH^{\theta/z}}{(2 n (z-1)+1) (-\theta +2 n (z-1)+1)}\Big(\frac{\delta}{H}\Big)^{\frac{2 n+\theta-1}{z}-2n}+\cdots\Big].
\end{align}
where the logarithmic term appears when $\theta=1-2\ell(1-z)$ for some integer $\ell$ and can be fitted with the NEC's. Note that there are logarithmic and double logarithmic terms for  $z=(2k-1)/(2k)$ and $\theta=0$. However, these are not consistent with the NEC's and we discard them.

\subsubsection{$z<0$}\label{zlt0}
In this case, the solution to \eqref{EoM-phi} read as
\begin{align}
&\varphi=\varphi_0+\varphi_{2}h^{2}+\varphi_{4}h^{4}+\varphi_{6}h^{6}+\tilde{\varphi}_1h^{4-2z}+\cdots,
\end{align}
where coefficients $\varphi_{2n}$ are functions of $z$ and $\theta$. However, near the boundary, only the constant term $\varphi_0$ contributes to the singular part of the integrand of \ref{int 1} as,
\begin{equation}\label{eq 4}
\mathcal{L}_h \sim -\frac{\sin  \varphi_{0}}{h^{2}} +\cdots,
\end{equation}
Hence, by adding and subtracting the singular term \eqref{eq 4}, we can isolate the divergent part of the entropy, and  make it finite as below
\begin{align}
S_{EE}&=\frac{L^{2}}{2G_{N}}\frac{1}{2z}\int_{\delta}^{r_{m}}\frac{dr}{r^{1-\frac{\theta}{z}}}\int_{h_{0}}^{h_{c}} dh\mathcal{L}_h 
\nonumber\\
&=\frac{L^{2}}{2G_{N}}\frac{1}{2z}(I_{1}+I_{2})
\end{align}
where
\begin{align}
I_{1}&=\int_{\delta}^{r_{m}}\frac{dr}{r^{1-\frac{\theta}{z}}}\int_{h_{0}}^{h_{c}} dh
\Big[\mathcal{L}_h +\frac{\sin  \varphi_{0}}{h^{2}} \Big]\,,\qquad
I_{2}=-\int_{\delta}^{r_{m}}\frac{dr}{r^{1-\frac{\theta}{z}}}\int_{h_{0}}^{h_{c}} dh
\frac{\sin  \varphi_{0}}{h^{2}} 
\end{align}
Now, we differentiate $I_1$ and $I_2$ with respect to $UV$ cut-off $\delta$ and look for various divergent terms,
\begin{align}
\frac{dI_{1}}{d\delta}&=-\frac{1}{\delta^{1-\frac{\theta}{z}}}\int_{h_{0}}^{0} dh
\Big[\mathcal{L}_h +\frac{\sin  \varphi_{0}}{h^{2}} \Big]+\cdots \,, \\
& \frac{dI_{2}}{d\delta}=\sin\varphi_{0}\Big[   \frac{1-\theta}{\theta}\frac{H^{\frac{1}{z}}}{\delta^{1-\frac{\theta}{z}+\frac{1}{z}}}  -\frac{r_m^{\frac{\theta}{z}}H^{\frac{1}{z}}}{\delta^{1+\frac{1}{z}}}+\frac{1}{\delta^{1-\frac{\theta}{z}}}\frac{1}{h_{0}}+\cdots\Big].
\end{align}
Note that as we see, at values $\theta=1$ and  $\theta=0$,  logarithmic terms appear in the expression of the entanglement entropy. But these are not compatible with the NEC when $z<0$. So we have 
\begin{align}
S_{EE}^{(y,z<0)}=\frac{L^{2}}{4G_{N}}\Big[\frac{\sin \varphi_{0}}{\theta-1}\Big(\frac{\delta}{H}\Big)^{\frac{\theta-1}{z}}+\Big(\frac{\sin \varphi_{0}}{h_{0}}-\int_{h_{0}}^{0} dh
\Big(\mathcal{L}_h +\frac{\sin  \varphi_{0}}{h^{2}} \Big)\Big)\frac{1}{\theta} (\frac{\delta}{H})^{\theta/z}\Big]
\end{align}
where the first and second terms are divergent when $\theta>1$ and $\theta>0$, respectively.


\section{Rotating the Entanglement Region}\label{sec:rotating}

In subsection \ref{subsec:shape-invariant}, we study the entanglement entropy for a singular region which has a bisector in the direction of $y$-axis. Now we want to rotate the region by $\pi/2$ angle such that to be in the direction of $x$-axis. To avoid repetition of calculation, we consider the following transformations which interchange the role of $x$ and $y$ coordinates. Let us consider new coordinates as 
\begin{align}
\tilde{r}=r^{1/z},\qquad \tilde{x}=x/z, \qquad \tilde{y}=y/z
\end{align}
so the metric in \eqref{metric1} changes to
\begin{align}\label{metric-tilde}
d\tilde{s}=\tilde{L}^2\tilde{r}^\frac{\tilde{\theta}}{\tilde{z}}\Big( \frac{d\tilde{r}^2+d\tilde{y}^2}{\tilde{r}^2}+\frac{d\tilde{x}^2}{\tilde{r}^{2/\tilde{z}}}\Big).
\end{align}
where $\tilde{L}=zL$, $\tilde{\theta}=\theta/z$ and $\tilde{z}=1/z$. The background \eqref{metric-tilde} is equivalent to \eqref{metric1} with interchanging the role of $x$ and $y$. So to get the entanglement entropy for this region, we can apply the following transformations to our previous results, 
\begin{align}\label{rescale}
L\rightarrow \tilde{L}=zL,\qquad z\rightarrow \tilde{z}=1/z, \qquad \theta\rightarrow \tilde{\theta}=\theta/z. 
\end{align}
In addition, we need to replace $(\delta,H,\varphi_0)\rightarrow(\delta^{z},H^{z},\pi/2-\varphi_0)$. The resulting entanglement entropy is found as follows.
\begin{enumerate}
	\item $z>1$:\\
	This case is derived from transforming results of subsection (\ref{zlt1}). We find,
	\begin{align}
	\tilde{\mathcal{L}}_h &\sim \sum_{n=0}\tilde{a}_n h^{2n(1-1/z)-2}+\cdots,\nonumber\\
	&\sim \frac{2\cos\varphi_{0}}{zh^{2}} + \frac{\sin \varphi_{0}\tan\varphi_{0}}{2z^2h^{2/z}}- \frac{\sin \varphi_{0}\tan^3\varphi_{0}}{8z^4h^{4/z-2}}+ \cdots.
	\end{align}
	and
	\begin{align}\label{rotate1}
	&S_{EE}^{(y,0<z<1)}\rightarrow S_{EE}^{(x,z>1)}\nonumber\\
	&=\frac{L^{2}}{2G_{N}}\frac{z^3}{2}\Big[\Big(\tilde{F}(h_0)-\int_{h_{0}}^{0} dh
	(\tilde{\mathcal{L}}_h  - \sum_{n=0}^{n_{max}}\tilde{a}_n h^{2n(1-1/z)-2})
	\Big)\frac{1}{\theta}\delta^{\theta/z} \nonumber\\
	&\frac{\tilde{a}_mH^{\theta/z}}{\theta}\log(\frac{\delta}{H})\nonumber\\
	&+\sum_{n=0}^{n_{max}}\frac{\tilde{a}_n H^{z\theta}}{(2 n (1-z)+z) (-\theta +2 n (1-z)+z)}\Big(\frac{\delta}{H}\Big)^{(2n-1)+\frac{\theta-2n}{z}}+\cdots\Big].
	\end{align}
	where the log term appears when $\theta=z-2m(z-1)$ for some integer $m$, the same as \eqref{thetam}. Again, putting $m=1$, an interesting special case is $\theta=0$ and $z=2$.  
	
	\item $0<z<1$:\\
	In this case we have
	\begin{align}
	\tilde{\mathcal{L}}_h &\sim \sum_{n=0}\tilde{b}_n h^{(2n(1-z)-(1+z))/z}+\cdots,\nonumber\\
	&\sim \frac{2\sin \varphi_{0}}{z^2h^{1+1/z}} + \frac{\cos \varphi_{0}\cot\varphi_{0}}{h^{3-1/z}}- \frac{z^2\cos \varphi_{0}\cot^3\varphi_{0}}{4h^{-5+3/z}}+\frac{z^4 \cos \varphi_{0}\cot^5\varphi_{0}}{8h^{-7+5/z}} + \cdots,
	\end{align}
	and	
	\begin{align}\label{rotate2}
	&S_{EE}^{(y,z>1)}\rightarrow S_{EE}^{(x,0<z<1)}\nonumber\\
	&=\frac{L^{2}}{2G_{N}}\frac{z^3}{2}\Big[\Big(F(h_0)-\int_{h_{0}}^{0} dh
	(\mathcal{L}_h  - \sum_{n=0}^{n_{max}}\tilde{b}_n h^{(2n(1-z)-(1+z))/z})
	\Big)\Big(\frac{1}{\theta}\delta^{z\theta}\Big)\nonumber\\
	&+\frac{\tilde{b}_mz^2H^{z\theta}}{\theta}\log (\frac{\delta}{H})
	\nonumber\\
	&+\sum_{n=0}^{n_{max}}\frac{\tilde{b}_n z H^{\theta/z} }{(1-2 n (1-z)) (2 n (1-z)+\theta-1)}\Big(\frac{\delta}{H}\Big)^{\frac{2 nz (1-z)+z\theta-z}{z}}+\cdots\Big].
	\end{align}
	where the log term exists when $\theta=z-2m(z-1)$ for some integer $m$. We discarded some log and double log terms which are not compatible with the NECs.
	
	\item $z<0$:\\
	\begin{align}\label{rotate3}
	&S_{EE}^{(y,z<0)}\rightarrow S_{EE}^{(x,z<0)}\nonumber\\
	&=\frac{L^{2}}{2G_{N}}\frac{z^3}{2}\Big(\frac{z\cos \varphi_{0}}{\theta-z}\Big(\frac{\delta}{H}\Big)^{z(\theta-z)}+(-\int_{h_{0}}^{0} dh
	\Big[\mathcal{L}_h +\frac{\cos  \varphi_{0}}{h^{2}} \Big]+\frac{\cos \varphi_{0}}{h_{0}})\frac{1}{\theta} (\frac{\delta}{H})^{z\theta}\Big)
	\end{align}
	
\end{enumerate}


\section{Conclusion}

In this paper, we studied the holographic entanglement entropy of anisotropic and nonconformal strongly coupled gauge theories, and explored the effects of singularity as well as spatial anisotropy on the entanglement entropy.

Our motivation was finding the universal terms just like in the case of the corner contribution in the 2+1-dimensions \cite{Ref44,Ref45} or higher dimensions with hyperscaling violation parameters \cite{Ref68} to the entanglement entropy in the isotropic spacetime.

We have shown that with respect to values of the $z$ and $\theta$, the new divergent structure may appear in the holographic entanglement entropy of the curved kink region. We identified these values of $z$ and $\theta$ that give rise to a universal contribution. Specially, we have found that for some values of $z$ and $\theta$, the logarithmic and double logarithmic terms appear. Unlike the homogeneous space-time in which the logarithmic term appears in the special value of $\theta=1$ which is reflected in the fact that the corresponding background provides a gravitational dual for the boundary theory with Fermi surface, in anisotropic space-time for some discrete values of $z$ and $\theta$ the logarithmic or double logarithmic term appears. In this case, we also found the area law violation.

To find explicit results, we investigated a shape invariant corner in an anisotropic space. We considered the parallel and transverse directions, i.e., singular surface oriented along the anisotropic scaling direction $y$ and laying along isotropic direction $x$, respectively.  

In the case of a singular entangling surface along the anisotropic direction, concerning various ranges of $z$, we found new universal contributions. For the case $z>1$, we found that for the values of $z=2k/(2k-1)$ with $k$ some positive integer and $\theta=0$, which corresponds to Lifshitz geometry, the double logarithmic term appears.
On the other hand, for the values of $\theta=z-2n(z-1)$, the logarithmic term appears. These ranges of $z$ and $\theta$ are consistent with the NEC's. Other than those values we found only power-law divergences. Note, for $z=1$, only power-law divergences appear \cite{Ref66}.

In the range $0<z<1$, our computations showed that, as before, for values $z=(2k-1)/(2k)$ and $\theta=0$ the logarithmic and double logarithmic terms appear, but these values are not consistent with the NEC's. On the other hand, taking $\theta_{\ell}=1-2\ell(1-z)$ with $\ell$ to be some positive integer leads to a logarithmic term and is consistent with the NEC's. For other values, we have power-law divergences.  

For the case $z<0$, we have not found any logarithmic term consistent with NEC's, and only power-law divergences appear.

\begin{table}[ht]
	\begin{center}
		\begin{minipage}{0.85\textwidth}
			\renewcommand{\arraystretch}{1.5}
			\begin{tabular}{|l| l| l| l|  l| l|}
				\hline
				Orientation	&    Parameters      				& $z > 1 $	     	& $0<z<1$		& $z<0$		\\
				\hline 
				Along $y$	&  $\theta=0$   and $z=2k/(2k-1)$   	& $\log^2\delta$  	&   power law    	&    power law 	\\
				&  $\theta=z-2m(z-1)$		        	&  $\log\delta$ 	&   power law          	&    power law	\\
				&   $\theta=1-2\ell(1-z)$  			&  power law  	&  $\log\delta$ 	&    power law 	\\
				\hline 
				Along $x$	&  $\theta=0$   and $z=2k/(2k-1)$   	& $\log^2\delta$  	&   power law    	&    power law 	\\
				&  $\theta=z-2m(z-1)$		        	&  $\log\delta$ 	&  $\log\delta$ 	&    power law 	\\
				\hline
			\end{tabular}
			\renewcommand{\arraystretch}{1}
			\caption{Summary of results for $\frac{2G_N}{L^2}S_{EE}$. We only consider cases compatible with the NEC's.  $m$ and $\ell$ are any positive integers. }\label{table:questions}
		\end{minipage}
	\end{center}
\end{table}
The summary of results are depicted in table \ref{table:questions}. 

The contributions of logarithmic or double logarithmic terms are universal in the sense that their values are independent of the details of the $UV$ regulator, so the appearance of that kind of terms helps us to probe the characteristics of the underlying theory. Note that, in
the cases that both logarithmic and double logarithmic terms appear simultaneously, the universal term is the double logarithmic term which is independent of the regularization scheme \cite{Ref51, Ref65}.

Here, it is worth commenting on properties in \eqref{a-properties} which are derived from subadditivity and Lorentz invariance. However, for the anisotropic space in subsection \ref{subsec:shape-invariant}, these properties \eqref{a-properties} can not be applied, due to the lack of rotational symmetry. In the asymmetric space, the entanglement entropy is not only a function of the opening angle, but also depends on the orientation of region. So it is not possible to derive the properties \eqref{a-properties} from adjacent regions.

By the way, we can investigate the smooth limit.
For the case of the region along $y$-axis, recall that in our notation $\Omega=\pi-2\varphi_0$, let us consider the limit 
$\varphi_0\rightarrow 0$ where the region approaches the half plane and corresponds to a smooth region. It is then expected that the coefficient of the universal term in \eqref{total-zgt1} vanishes, $a_k(0)=0$. On the other hand, for a pure state, the entanglement entropy of a region and its complement is the same, i.e. $S(V)=S(\overline{V})$. So we expect to have a Taylor expansion for $a_k(\varphi_0)$ with even powers of $\varphi_0$. This is indeed the case for $z>1$:
\begin{align}
a_k(\varphi_0)\sim \sin\varphi_0\tan^{2k-1}\varphi_0\simeq\varphi_0^{2k}\Big(1+\Big(\frac{2k}{3}-\frac{1}{2}\Big)\varphi_0^2+\cdots\Big).
\end{align}
This limit is not satisfied for $z<1$, where the universal term, $b_\ell\sim \cos\varphi_0\cot^{2\ell-1}\varphi_0$ diverges as $\varphi_0\rightarrow 0$. This is because of the breakdown of our expansion in \eqref{L-zlt1} for small $\varphi_0$ where coefficients are very large.    


Finally, it would be interesting to investigate these divergence structures in higher dimensions for various kinds of singularities or smooth cases in general dimensions.\\

\section*{Acknowledgment}
We would like to thank Dimitrios Giataganas and Juan F. Pedraza for useful comments and discussions.  
MG would like to thank Sepideh Mohammadi for providing encouragement and helpful comments. This work was fully supported by people of Iran.

\appendix
\section{Solving the equation of motion}\label{app1}
By extremizing the entropy functional \eqref{SEE-1} we derive the equation of motion for $r(\rho,\varphi)$,
\begin{align}\label{EoMr}
&2 \rho ^{2 b} (\theta -z-1) ((b-1) \cos 2 \varphi -b-1) r^2 \Big(8 b^2 \dot{r}^2 \sin ^2\varphi   \nonumber\\
&+\rho ^2 \Big(8 r'^2 \cos ^2\varphi +2 (-(b-1) \cos 2 \varphi +b+1)^2\Big)-8 b \rho  \dot{r} r' \sin 2 \varphi \Big) \nonumber\\
&+4 z \rho ^{2 b} r^3 \Big[-4 b^2 \sin ^2\varphi  ((b-1) \cos 2 \varphi -b-1) \ddot{r} \nonumber\\
&+2 b \sin 2 \varphi  \Big(-b^2+(b-1)^2 \cos 2 \varphi +4 b+1\Big) \dot{r}+\rho  \Big(4 b \sin ^2\varphi  ((b-1) \cos 2 \varphi +3 b-1) r'\nonumber\\
&+4 \cos \varphi  ((b-1) \cos 2 \varphi -b-1) \Big(2 b \dot{r}' \sin \varphi -\rho r'' \cos \varphi  \Big)\Big)\Big]  \nonumber\\
&+4 r^{2/z} \Big[z r \Big(\rho ^2 \dot{r} \Big(4 (b-1) \sin 2 \varphi  r'^2+8 ((b-1) \cos 2 \varphi -b-1) \dot{r}' r' \nonumber\\
&+2 \sin 2 \varphi  \Big(-b^2+(b-1)^2 \cos 2 \varphi -4 b+1\Big)\Big)+\rho ^2 \Big(\rho  \Big(\Big(4 \Big(b^2-b+1\Big) \cos 2 \varphi  \nonumber\\
&-\Big(b^2-3 b+2\Big) \cos 4 \varphi  -3 b^2+b+6\Big) r'-4 \sin \varphi  ((b-1) \cos 2 \varphi -b-1) \Big(\rho  r'' \sin \varphi  \nonumber\\
&+2 \dot{r}' \cos \varphi \Big)+8 r'^3\Big)-2 ((b-1) \cos 2 \varphi -b-1) \ddot{r} \Big(2 r'^2+\cos 2 \varphi +1\Big)\Big)\nonumber\\
&-4 (b-1) b \dot{r}^3\sin 2 \varphi  +4 \rho  \dot{r}^2 \Big(b ((b-1) \cos 2 \varphi -b+5) r'+\rho  (-(b-1) \cos 2 \varphi +b+1) r''\Big)\Big) \nonumber\\
&+4 \rho ^2 (\theta -2) ((b-1) \cos 2 \varphi -b-1) \Big(\rho  r' \sin \varphi +\dot{r}\cos \varphi  \Big)^2\Big]=0
\end{align}

Solving the above equation in the general case is horrible. So we consider an important  special case by setting $b=1/z$.  This corresponds to a shape invariant region under scaling (\ref{scaling}) and greatly simplifies the EoM (\ref{EoMr}).

Setting $b=1/z$, we then adopt the ansatz $r=\rho h^z(\varphi)$. The boundary condition would be $h=0$ as $\varphi \rightarrow \varphi_0$. 
This ansatz simplifies (\ref{EoMr}) to an ordinary nonlinear differential equation for $h(\varphi)$, 
\begin{align}\label{eq-h}
&4z^{2}(1+z+(-1+z)\cos 2\varphi) h\Big(z^{2}\cos ^{2}\varphi+h^{2z}(z^{2}+\frac{\sin ^{2}\varphi}{h^{2}})\Big)\ddot{h}+4(-1+z)z^{5}h^{-1+2z}\sin 2\varphi\dot{h}^{3}
\nonumber\\
&+z^{2}\Big(4z^{2}(1+z-\theta)\cos ^{2}\varphi(1+z+(-1+z)\cos 2\varphi)
\nonumber\\
&+4h^{2z}(-(-1+z)z^{2}(-2+z+(2+z)\cos 2\varphi)+(2z-\theta)(1+z+(-1+z)\cos 2\varphi)\frac{\sin ^{2}\varphi}{h^{2}})\Big)\dot{h}^{2}
\nonumber\\
&+2z^{2}h\Big(z(3+2z+3z^{2}-2(1+z)\theta+(-1+z)(3+3z-2\theta)\cos 2\varphi)
\nonumber\\
&-(2(-1+z)z^{2}+\frac{1}{h^{2}}(3+2z+3z^{2}-2(1+z)\theta+(-1+z)(3+3z-2\theta)\cos 2\varphi))h^{2z}\Big)\sin 2\varphi\dot{h}
\nonumber\\
&+h^{2}\Big(z^{2}(3+z(3+6z-\theta)-3\theta+4(-1+(-1+z)z+\theta)\cos 2\varphi-(-1+z)(1+2z-\theta)\cos 4\varphi)
\nonumber\\
&+\frac{1}{h^{2}}(1+z-\theta)(1+z+(-1+z)\cos 2\varphi)^{3}
+z^{2}h^{2z}[8z^{2}+\frac{1}{h^{2}}(6+3z(1+z-\theta)-\theta
\nonumber\\
&+4(-1+z+z^{2}-z\theta)\cos 2\varphi+(-1+z)(2+z-\theta)\cos 4\varphi)]\Big)
=0.
\end{align}
Now we change the independent variable from $\varphi$ to $h$. Using the relations $\dot{h}=\frac{1}{\dot{\varphi}(h)}$, $\ddot{h}=-\frac{\ddot{\varphi}}{\dot{\varphi}^{3}}$ in \eqref{eq-h}, we reach to the following equation of motion for $\varphi(h)$, 
\begin{align}\label{EoM-phi}
&-4z^{2}h(1+z+(-1+z)\cos 2\varphi)\Big(z^{2}\cos ^{2}\varphi+h^{2z}(z^{2}+\frac{\sin ^{2}\varphi}{h^{2}})\Big)\ddot{\varphi}
\nonumber\\
&+h^{2}\Big(z^{2}(3+z(3+6z-\theta)-3\theta+4(-1+(-1+z)z+\theta)\cos 2\varphi
\nonumber\\
&-(-1+z)(1+2z-\theta)\cos 4\varphi)
+(1+z-\theta)(1+z+(-1+z)\cos 2\varphi)^{3}\frac{1}{h^{2}}
\nonumber\\
&+z^{2}h^{2z}(8z^{2}+(6+3z(1+z-\theta)-\theta+4(-1+z+z^{2}-z\theta)\cos 2\varphi
\nonumber\\
&+(-1+z)(2+z-\theta)\cos 4\varphi)\frac{1}{h^{2}})\Big)\dot{\varphi}^{3}
\nonumber\\
&+2z^{2}h\sin 2\varphi\Big(z(3+2z+3z^{2}-2(1+z)\theta+(-1+z)(3+3z-2\theta)\cos 2\varphi)
\nonumber\\
&-h^{2z}(2(-1+z)z^{2}+(3+2z+3z^{2}-2(1+z)\theta+(-1+z)(3+3z-2\theta)\cos 2\varphi)\frac{1}{h^{2}})\Big)\dot{\varphi}^{2}
\nonumber\\
&+\Big(4z^{4}(1+z-\theta)\cos ^{2}\varphi(1+z+(-1+z)\cos 2\varphi)
\nonumber\\
&+z^{2}h^{2z}(-4(-1+z)z^{2}(-2+z+(2+z)\cos 2\varphi)
\nonumber\\
&+4(2z-\theta)(1+z+(-1+z)\cos 2\varphi)\frac{\sin ^{2}\varphi}{h^{2}})\Big)\dot{\varphi}
+4(-1+z)z^{5}h^{2z-1}\sin 2\varphi=0
\end{align}
where $\dot{\varphi}(h)=\frac{d\varphi}{dh}$, $\ddot{\varphi}(h)=\frac{d^{2}\varphi}{dh^{2}}$.  
Now, we solve this equation perturbatively in terms of $h$ near the boundary, where $h$ is small. The boundary condition read as $\varphi(0)=\varphi_{0}$ at $h=0$. Let us consider different $z$ ranges as follows.

{\bf a) $\mathbf{z>1}$}\\
For $z>1$, we find the following expansion as a solution to \eqref{EoM-phi},
\begin{align}
&\varphi=\varphi_0+\sum_{n=1}\varphi_{2n}(h^{2})^{n(z-1)+1}
\nonumber\\
&\quad=\varphi_0+\varphi_{2}h^{2z}+\varphi_{4}(h^{2})^{2z-1}+\varphi_{6}(h^{2})^{3z-2}+\cdots,
\end{align}
where some $\varphi_{2n}$'s are derived as,
\begin{align}
\varphi_2&=\frac{(z-1) \tan \varphi_0}{(\theta+z-2) ((z-1) \cos (2\varphi_0)+z+1)},\nonumber\\
\varphi_4&=-\frac{(\theta-1) \tan ^2\varphi_0}{z (2 z-1) (\theta+3 z-4)}\varphi_2 \nonumber
\end{align}

{\bf b) $\mathbf{0<z<1}$}\\
Here, from \eqref{EoM-phi} we find,
\begin{align}
&\varphi=\varphi_0+\varphi_{2}h^{2}+\varphi_{4}h^4+\tilde{\varphi}_1h^{4-2z}+\tilde{\varphi}_{3}^{6-4z}+\cdots,
\end{align}

{\bf c) $\mathbf{z<0}$}\\
In this case, the solution to \eqref{EoM-phi}
read as
\begin{align}
&\varphi=\varphi_0+\varphi_{2}h^{2}+\varphi_{4}h^{4}+\varphi_{6}h^{6}+\tilde{\varphi}_1h^{4-2z}+\cdots,
\end{align}
where coefficients $\varphi_{2n}$ are functions of $z$ and $\theta$.


\end{document}